\documentclass[aps,prb,showpacs,superscriptaddress,twocolumn]{revtex4}
\usepackage{times}
\usepackage{color}
\usepackage{graphicx,epsfig}
\usepackage{dcolumn}
\usepackage{amsmath}
\usepackage{amssymb}
\usepackage{epstopdf}
\usepackage{amsmath,verbatim}
\usepackage{bm}

\def\kv{{\bf k}}
\def\qv{{\bf q}}

\def\ev{{\bf e}} 

\def\nn{\nonumber}
\begin{document}

\title{Theory  of ultrafast  quasiparticle dynamics  in  high-temperature 
superconductors: Pump fluence dependence}

\author{Jianmin Tao}
\altaffiliation{Current address: Department of Physics,
Tulane University, New Orleans, Louisiana 70118}
\affiliation{Theoretical Division and Center for Nonlinear Studies,
Los Alamos National Laboratory, Los Alamos, New Mexico 87545}

\author{Rohit P. Prasankumar}
\affiliation{Center for Integrated Nanotechnologies, Los Alamos 
National Laboratory, Los Alamos, New Mexico 87545}

\author{Elbert E. M. Chia}
\affiliation{Division of Physics and Applied Physics, School of
Physical and Mathematical Sciences, Nanyang Technological
University, Singapore 637371, Singapore}

\author{Antoinette J. Taylor}
\affiliation{Center for Integrated Nanotechnologies, Los Alamos 
National Laboratory, Los Alamos, New Mexico 87545}

\author{Jian-Xin Zhu}
\email[To whom correspondence should be addressed. \\ ]{jxzhu@lanl.gov}
\homepage{http://theory.lanl.gov}
\affiliation{Theoretical Division and Center for Nonlinear Studies,
Los Alamos National Laboratory, Los Alamos, New Mexico 87545}

\date{\today}
\begin{abstract}
We present a theory for the time-resolved optical spectroscopy of high-temperature superconductors at high excitation densities with strongly anisotropic electron-phonon coupling. A signature of the strong coupling between the out-of-plane, out-of-phase O buckling mode ($B_{1g}$) and electronic states near the antinode is observed as a higher-energy peak in the time-resolved optical conductivity and Raman spectra, while no evidence of the strong coupling between the in-plane Cu-O breathing mode and nodal electronic states is observed.  More interestingly, it is observed that under appropriate conditions of pump fluence, this signature exhibits a re-entrant behavior with time delay, following the fate of the superconducting condensate.  
\end{abstract}
\maketitle

\section{Introduction} 

Since its discovery in 1986, high-temperature superconductivity in cuprates has 
been a central topic of study in condensed matter physics. It is now widely believed 
that Cooper pair formation is essential for the superconducting condensate in these 
systems. However, the nature of the mediator (or glue) responsible for Cooper pairing remains 
hotly debated. Although the interaction of electrons with  lattice vibrations is not likely to solely account for the essential properties of high-$T_c$ superconductors (HTSCs), many probes including angle-resolved photoemission,~\cite{ALanzara:2001,TCuk:2004,
TPDevereaux:2004,GHGweon:2004,WMeevasana:2006} inelastic neutron 
scattering,~\cite{DReznik:2006}, tunneling~\cite{JLee:2006,JXZhu:2006a} and 
Raman~\cite{MOpel:1999} spectroscopies have revealed 
that electron-phonon interactions have significant effects on various properties.
Complementary to the time-integrated techniques, different ultrafast pump-probe 
techniques~\cite{JDemsar:1999,PKusar:2008,RDAveritt:2002,EMChia:2011,RAKaindl:2000,
RPSaichu:2010} have been used to disentangle microscopic interactions 
in HTSCs. These techniques aim to study the recombination of 
photoexcited quasiparticles and the resulting recovery of the superconducting condensate. In HTSCs, 
time-resolved (TR) angle-resolved photoemission spectroscopy~\cite{LPerfetti:2007} and 
TR optical reflectivity~\cite{NGedik:2005} have indicated that the excited quasiparticles 
 preferentially couple to a small number of phonon subsets before decaying 
through anharmonic coupling to all other lattice vibrations, in support of the notion 
that selective optical phonon modes give rise to anisotropy of the 
electron-phonon (el-ph) coupling.~\cite{TPDevereaux:2004} In addition, this anisotropy
has also been observed in TR electron 
diffraction,~\cite{FCarbone:2008} while resonant femtosecond study of 
both electronic and phononic degrees of freedom 
suggests strong el-ph coupling.~\cite{APashkin:2010}

Despite considerable progress in pump-probe experimental studies, work on
microscopic modeling of the influence of el-ph interactions on observables, 
such as the time-dependent optical conductivity or Raman spectra, is very limited.
Understanding the non-equilibrium dynamics of quantum many-body systems 
has, in fact, posed a theoretical challenge. Historically, theoretical attempts 
to model the time evolution of properties 
have either used quasi-equilibrium models such as $T^{*}$ and $\mu^{*}$ 
models~\cite{EJNicol:2003} 
to describe non-equilibrium excitations created by a pump pulse~\cite{VVKabanov:1999} 
or rate equation approaches based on the phenomenological Rothwarf-Taylor 
model~\cite{ARothwarf:1967} to describe the recovery dynamics of the superconducting state. 
Recently, the time evolution of the optical conductivity has been studied within a 
microscopic model that treats the excitation and relaxation dynamics on the same 
footing.~\cite{JUnterh:2008} All these theories are suitable for pump-probe 
experiments with low excitation fluence,
where the superconducting condensate is merely perturbed but not destroyed.  
A picture of the dynamics of quasiparticles and the superconducting condensate in the 
photo-induced phase transition regime,~\cite{CGiannetti:2009,MBeyer:2011,GLDakovski:2011}
 which is impulsively driven by a high excitation fluence, has as yet been beyond reach.

Here we formulate a theory for the TR optical conductivity and TR Raman scattering in 
HTSCs in the regime of intermediate to high intensity of pump fluence. The theory is aimed to address directly 
the situation where the superconducting condensate can be destroyed by pump pulse. 
It  is based on an effective temperature model
for different subsystems contributing to the response:  electrons, 
hot phonons (i.e., out-of-plane out-of-phase O buckling $B_{1g}$ phonons and 
half-breathing in-plane Cu-O bond  stretching phonons) that are strongly coupled to electrons, 
and the cold lattice. The model phenomenologically includes the effect of the pump pulse  
but addresses in greater depth the electron-hot phonon coupling based 
on a microscopic model Hamiltonian
for $d$-wave superconductivity in HTSCs. This microscopic treatment goes beyond 
previous effective 
models for the normal state,~\cite{PBAllen:1987,LPerfetti:2007} allowing us to describe the
quasiparticle dynamics in both the normal and superconducting states with the same approach. 
Within this unified model, the time evolution of the whole set of experimental measurables 
can be calculated in a streamlined way. Our first test of this approach considered the 
$B_{1g}$ phonons as the only hot phonon mode in the calculation of the TR spectral 
function for a very high pump fluence.~\cite{JTao:2010} 
In the present work, we include the half-breathing  
phonons as a second hot phonon mode in the calculation of the TR optical spectroscopy. 
Importantly, the influence of excitation density on quasiparticle dynamics in HTSCs is also
investigated. Our  calculations show that, in the superconducting 
state, in addition to the peak in the optical conductivity and Raman spectra due to 
the Drude response, there is another peak at 
higher frequencies. This high frequency peak disappears when the system
evolves into the normal state, but recurs if the superconducting condensate is recovered, 
suggesting the significance
of the superconducting gap in the TR optical properties. 

The outline of the paper is as follows: In Sec.~\ref{sec:temp}, we 
lay down  the effective-temperature model for the $d$-wave superconductor with electronic coupling to both $B_{1g}$ and half-breathing stretching phonon modes. The time-dependent effective temperatures for the respective subsystems are evaluated depending on the strength of pump fluence. With the obtained time-dependence of effective temperatures, the time-resolved optical conductivity and Raman spectra and their pump-fluence dependence are presented in Secs.~\ref{sec:sigma} and \ref{sec:chi}.  Finally, a conduction is given in Sec.~\ref{sec:conclusion}.

\section{Effective temperature model}
\label{sec:temp} 

Let us consider a two-dimensional superconductor exposed to a laser field.
The model Hamiltonian can be written as~\cite{JUnterh:2008,JTao:2010}
\begin{eqnarray}\label{hamiltonian}
H &=& \sum_{\kv\sigma}\xi_{\kv}c^{\dagger}_{\kv\sigma}
c_{\kv\sigma}+
\sum_{\kv}(\Delta_{\kv} c_{\kv\uparrow}^{\dagger}
 c_{-\kv\downarrow}^{\dagger}+{\rm h.c.}) 
+
\sum_{\qv\nu}\hbar\Omega_{\nu\qv} \nn \\
&\times&
\bigg(b_{\nu\qv}^{\dagger} b_{\nu\qv}+\frac{1}{2}\bigg)
+
\frac{1}{\sqrt{N_L}}\sum_{\kv\qv\nu\sigma}g_\nu(\kv,\qv)
c^{\dagger}_{\kv+\qv,\sigma} c_{\kv\sigma} A_{\nu\qv} \nn \\
&+& H_{\rm field}(\tau),
\end{eqnarray}
where $c^{\dagger}_{\kv\sigma}$ ($b_{\nu\qv}^{\dagger}$) and $c_{\kv\sigma}$ 
($b_{\nu\qv}$) are the creation and annihilation operators for an electron 
with momentum $\mathbf{k}$ and
spin $\sigma$ (phonon with momentum $\mathbf{q}$ and vibrational mode 
$\nu$; $\nu=1,2$ represent the $B_{1g}$ and half-breathing modes, respectively), 
$ A_{\nu\qv} =  b_{\nu,-\qv}^{\dagger}+ b_{\nu\qv}$, 
$\xi_{\mathbf{k}}$  is the normal-state energy dispersion,
 $\mu$ the chemical potential, 
$\Delta_\kv=(\Delta_0/2)(\cos k_x - \cos k_y)$ is the $d_{x^2-y^2}$-wave gap 
function,  $N_{L}$ is the total number of lattice sites, 
and $g_\nu$ the coupling matrix.  Following the procedure sketched 
in Ref.~\onlinecite{JTao:2010}, we arrive at a four-temperature model:
\begin{subequations}
\begin{eqnarray}
\frac{\partial T_e}{\partial \tau} &=& \frac{1}{C_e}\sum_{\nu} K_{\nu}(T_{e},T_{ph,\nu}) 
+\frac{P_e}{C_e}\;, \label{three-Te} \\ 
\frac{\partial T_{ph,\nu}}{\partial \tau} &=& - \frac{K_{\nu}(T_{e},T_{ph,\nu})}{C_{ph,\nu}}
 -\frac{T_{ph,\nu}-T_l}{\tau_{\beta,\nu}}\;, 
\label{three-Tph} \\
\frac{\partial T_l}{\partial \tau} &=&
\sum_{\nu} \bigg(\frac{C_{ph,\nu}}{C_l}\bigg)\frac{T_{ph,\nu}-T_l}{\tau_{\beta,\nu}}\;.
\label{three-Tl}
\end{eqnarray}
\label{EQ:effectiveT}
\end{subequations}
Here $K_{\nu}$ is the el-ph coupling kernel, which can be calculated from
the model Hamiltonian~(\ref{hamiltonian}) with the equation-of-motion approach. It
is given by
\begin{eqnarray}\label{rate}
K_{\nu} &=&
\frac{4\pi}{N_L}\sum_{\kv\qv} g_\nu^2(u_\kv u_{\kv-\qv}-v_\kv v_{\kv-\qv})^2
\delta (E_{\kv-\qv}-E_\kv-\Omega_{\nu\qv}) \nn \\
&\times&
\Omega_{\nu\qv}  \bigg[e^{(\beta_{ph}-\beta_e)\Omega_{\nu\qv}}-1\bigg]
f_{\mathbf{k}}(1-f_{\mathbf{k}-\mathbf{q}})N_{\Omega_{\nu\qv}}\;,
\end{eqnarray} 
where the Bogoliubov amplitudes are
$u_{\kv} = [(1+\xi_\kv/E_\kv)/2]^{1/2}$ and $v_\kv = 
\text{sgn}(\Delta_{\kv})[(1-\xi_\kv/E_\kv)/2]^{1/2}$ with 
$E_\kv = \sqrt{\xi_\kv^2+\Delta_\kv^2}$ being the quasiparticle energy, and
the Fermi-Dirac and Bose-Einstein distribution functions are given by
$f_{\mathbf{k}} = f(E_\kv)=1/(e^{\beta_e E_\kv}+1)$, and $N_{\Omega_0}=
N(\Omega_0)=1/(e^{\beta_{ph} \Omega_0}-1)$ with $\beta_{e(ph,\nu)} = 1/k_B T_{e(ph,\nu)}$. 
In Eq.~(\ref{EQ:effectiveT}), the specific heat for electrons per unit cell  is found to be 
\begin{equation}
\label{eheat}
C_e =
\frac{\beta_e k_B}{N_{L}} \sum_\kv\bigg[-\frac{\partial f(E_\kv)}{\partial E_\kv}\bigg]\bigg(2E_\kv^2+
\beta_e \Delta_k \frac{\partial \Delta_k}{\partial \beta_e}\bigg)\;,
\end{equation}
while that for each hot phonon mode is given by
\begin{equation}
\label{phheat}
C_{ph,\nu} = \frac{k_B}{4}(\hbar \Omega_{\nu}\beta_{ph,\nu})^2
\bigg[{\rm coth}^2\bigg(\frac{\hbar \Omega_{\nu}\beta_{ph,\nu}}{2}\bigg)-1\bigg]\;
\end{equation}
in the Einstein mode approximation $\Omega_{\nu\qv} = \Omega_{\nu}$.
Finally, $P_e$ is the power intensity (i.e., power per unit cell) for pumping electrons 
and $\tau_{\beta\nu}$ the anharmonic decay time of each hot phonon mode.

\begin{figure}[t]
\includegraphics[width=0.8\columnwidth]{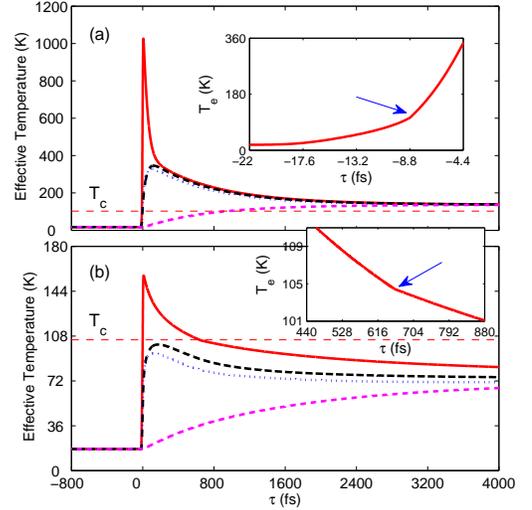}
\caption{(Color online) Time evolution of effective temperatures of electrons 
$T_e$ (red-solid line), 
$B_{1g}$-phonon mode $T_{ph,1}$ (blue-dotted line), half-breathing phonon mode 
$T_{ph,2}$ (black-dashed line), and cold lattice $T_{l}$ (magenta-dashed line)
for the powers (a) $P_0=0.436\;\mu\text{W}$ and (b) $P_0=0.022\;\mu\text{W}$. 
The inset in panel (a) shows a zoomed-in view of $T_e$ rising up above $T_c$, while 
that in panel (b) shows an enlarged view of $T_e$ cooling down below $T_c$.}
\label{FIG:EffectiveT}
\end{figure}

Throughout the paper, we use a five-parameter tight-binding model~\cite{MRNorman:1995}
to describe the normal-state energy dispersion, which is typical to the optimally doped 
Bi$_2$Sr$_2$CaCu$_2$O$_{8+x}$ (Bi-2212):
\begin{eqnarray}
\xi_{\mathbf{k}} &=& -2t (\cos k_x + \cos k_y) -4t^{\prime} \cos k_{x}
\cos k_y  \nonumber \\
&& -2t^{\prime\prime} (\cos 2k_x + \cos 2k_y) 
\nonumber \\
&&-4t^{\prime\prime\prime} (\cos 2k_x \cos k_y
+ \cos k_x \cos 2k_y)  \nonumber \\
&&
-4 t^{\prime\prime\prime\prime} \cos 2k_x \cos 2k_y - \mu\;, 
\end{eqnarray}
where  the hopping integrals
$t=1$, $t^{\prime}=-0.2749$,
$t^{\prime\prime}=0.0872$, $t^{\prime\prime\prime}=0.0938$, 
$t^{\prime\prime\prime\prime}=-0.0857$, and
$\mu=-0.8772$. The absolute energy of $t$ is 150 meV.
A feature of this dispersion is a flat band with a saddle point at the $M$ points of the Brillouin zone.
The temperature-dependence of the $d$-wave gap magnitude 
 is given by~\cite{FGross:1986} 
 \begin{equation}
\Delta_0(T_e) = \Delta_{00} {\rm tanh}\{(\pi/z)\sqrt{ar(T_c/T_e-1)}\}\;,
\end{equation}
where $z=\Delta_{00}/(k_BT_c)$. 
In our calculations, we
set $\Delta_{00} = 30\;\text{meV}$, the critical temperature $T_c = 104\;\text{K}$ (from the setting of $T_c=0.06t$ for simplicity), 
the specific heat jump at $T_c$ is $r=\Delta C_e/C_e \sim 1.43$,
and $a=2/3$.  We take the anisotropic el-ph coupling in the form given in Refs.~\onlinecite{TPDevereaux:2004,JXZhu:2006b} 
with $\Omega_{1} =45\;\text{meV} $ 
and $g_{1}^{(0)} = 90\;\text{meV}$ 
and $\Omega_{2} = 70\;\text{meV}$ 
and 
$g_{2}^{(0)} = 120\;\text{meV}$, 
$\tau_{\beta,1} =\tau_{\beta,2}=880\;\text{fs}$,  
and
$C_{ph,1} = C_{ph,2} = 0.2 C_l$. 
The pump is represented by a Gaussian pulse $P=P_0 e^{-\tau^2/(2\sigma^2)}$, with
a FWHM (full width at half maximum) of 2.35 $\sigma$. Hereafter, we set 
$\sigma = 4.4\;\text{fs}$, 
giving a FWHM of about 10.34 fs. This value is smaller than the commonly used 
experimental values of about 30-50 fs but is indeed close to that value of 12 fs used in the recent TR experiment on 
YBa$_2$Cu$_3$O$_{7-\delta}$.~\cite{APashkin:2010} We believe that this variation will not affect the qualitative physics, as will be 
presented below. 
We take the number of $\mathbf{k}$ points to be $40 \times 40$
in the Brillouin zone for the temperature evolution and use $256\times 256$ for the TR
optical conductivity and Raman scattering spectral function.  All calculations
are done with the system initially in the superconducting state, for $T\ll T_c$.

Figure~\ref{FIG:EffectiveT} shows the time evolution of the effective temperature 
for each subsystem for (a) a large pump power intensity $P_0=0.436\;\mu\text{W}$ 
and (b) an intermediate value of 
$P_0=0.022\;\mu\text{W}$. 
These values of pump power corresponds to $120\;\mu\text{J}/\text{cm}^2$ and $6\;\mu\text{J}/\text{cm}^2$ 
of pump fluence in Bi-2212 by assuming a 60 nm of optical absorption depth.  
Starting from the initial temperature $T_e=T_{ph,\nu}=T_{l}=17\;\text{K}$, 
the electron temperature $T_e$ increases rapidly after photoexcitation, and the superconductor is driven into the normal state while exhibiting a kink structure at $T_c$ (inset of Fig.~\ref{FIG:EffectiveT}(a)) for both power intensities. 
In addition, $T_e$ shows a very narrow peak for the large $P_0$ 
(see Fig.~\ref{FIG:EffectiveT}(a)), with a broader peak for the intermediate 
$P_0$ (see Fig.~\ref{FIG:EffectiveT}(b)), due to the fact  that the highest  temperature achieved by electrons
is very sensitive to the pump fluence. 
In both cases, the hot phonon subsystems are first heated up through their coupling to the photoexcited electrons, and after reaching their maximum temperature, they cool down by dissipating energy into the cold lattice through anharmonic coupling. A noticeable difference between the large and intermediate pump fluences is that in the latter case, the superconducting state recovers more rapidly (i.e.,  $T_e \leq T_c$) in a very short time ($\sim 650\;\text{fs}$), 
with the kink recurring during the cooling stage (inset of Fig.~\ref{FIG:EffectiveT}(b)). Further energy relaxation is then slowed down significantly due to the opening of the superconducting gap.  

\section{Time-resolved optical conductivity}  
\label{sec:sigma} 

Within the Kubo formalism, 
the real part of the TR optical conductivity is given by:~\cite{FMarsiglio:1991}
\begin{equation}
 \sigma_{1}(\omega) = -\frac{e^2 {\rm Im} \Pi (\omega)}{\omega} \;,
\end{equation}
where 
\begin{eqnarray}\label{impi}
{\rm Im} \Pi (\omega) = -\frac{2\pi^{2}}{N_L}\sum_\kv 
\biggl{(}\frac{\partial \xi_\kv}{\partial k_x}\biggr)^{2} I_\Pi(\kv,\omega)\;,
\end{eqnarray}
and
\begin{eqnarray}\label{integral}
I_\Pi(\kv,\omega) &=& 2i \int d\tau^{\prime} \text{Tr}\text{Im} [{\hat F}^*(\kv,\tau^{\prime}){\hat A}(\kv,\tau^{\prime})]
e^{i\omega \tau^{\prime}}\;.
\end{eqnarray}
Here ${\hat A}(\kv,\tau^{\prime})$ 
and ${\hat F}(\kv,\tau^{\prime})$ 
 are the Fourier transform of the TR spectral functions ${\hat A}(\kv,\epsilon)$ and 
 ${\hat F}(\kv,\epsilon) \equiv {\hat A}(\kv,\epsilon)f(\epsilon)$.   
In the derivation of Eq.~(\ref{impi}), we have used the Hilbert transform
\begin{equation}
\hat g(\kv,i\omega_n) = \int_{-\infty}^{\infty}d\epsilon \frac{{\hat A}(\kv,\epsilon)}
{i\omega_n-\epsilon}\;,
\end{equation}
 with the single-particle spectral function
\begin{equation}
{\hat A}(\kv,\epsilon) = -\frac{1}{\pi}~ 
{\rm Im} [{\hat g}(\kv,i\omega_n \rightarrow \epsilon+i\delta)]\;,
\end{equation} 
where the Green's function ${\hat g}$ and $\hat{A}$ are $2\times 2$ matrices 
in the Nambu space. Since this spectral function as calculated with the method 
of Ref.~\onlinecite{JTao:2010} is a function of the effective electronic 
temperature, which is time dependent (see the discussion in Sec.~\ref{sec:temp}),  
it is time resolved.  Therefore, the optical conductivity and the Raman 
spectra as discussed in the next section are also time dependent.  

\begin{figure}
\includegraphics[width=0.8\columnwidth]{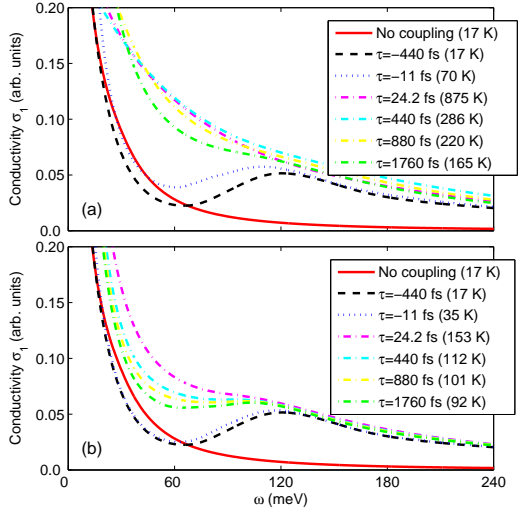}
\caption{(Color online) 
Real part of the time-resolved optical conductivity at
several different time delays for (a) $P_0=0.436\;\mu\text{W}$ and (b) $0.022\;\mu\text{W}$. 
The number in parentheses is $T_e$. 
}
\label{sigma}
\end{figure}

Figure~\ref{sigma} shows the time evolution of the real part 
of the optical conductivity 
$\sigma_1(\omega)$ at several
selected time delays.
From Fig.~\ref{sigma}, one can see that at all 
time delays, the optical conductivity shows the well-known Drude
peak at $\omega = 0$, due to 
the nodal quasiparticles for the 
$d$-wave gap symmetry.  In addition, for $\tau=-440$ fs and 
$-11$ fs, at which the material is superconducting and $\Delta_0 \approx 30$ meV, 
we observe that $\sigma_1(\omega)$ exhibits a broad peak at about 
$\omega = 2\Delta_0 + \Omega_{ph,1}$. (The specific location 
may be affected by several factors, although $2\Delta_0 + \Omega_{ph,1}$ plays
a substantial role.) Our observation is consistent with 
earlier study of optical conductivity  in the thermal equilibrium state of 
HTSCs.~\cite{SMaiti:2010}
In contrast, no peak at $\omega=\Omega_{ph,2}$ is observed 
(a signature of the coupling between electrons and the half-breathing 
mode along the nodal directions). 
This is due to the fact that  the 
coupling between electrons and the $B_{1g}$ phonon mode is the strongest at the
$M$ points, at which the van Hove singularity is also located.
This is further verified by the observation that no such peak appears when the 
el-ph coupling is turned off in the initial superconducting state
$T_e=17\;\text{K}$ (red solid line in Fig.~\ref{sigma}). 
After photoexcitation, the superconducting 
gap, with its time-dependence encoded in the effective electron temperature $\Delta_0(T_e)$,  
is decreased and the high-frequency peak shifts toward lower frequencies, merging into the 
zero-frequency Drude peak in the normal state. For large $P_0$, the Drude 
peak remains for the whole time-delay range being simulated.  However, 
for intermediate $P_0$, once the superconducting condensate recovers, 
the  ($2\Delta_0(T_e) + \Omega_{ph,1}$)-peak recurs--- a re-entrant behavior. 
The absence of the peak at $\Omega_{ph,2}$ is robust with time delay.

\begin{figure}[t]
\includegraphics[width=0.8\columnwidth]{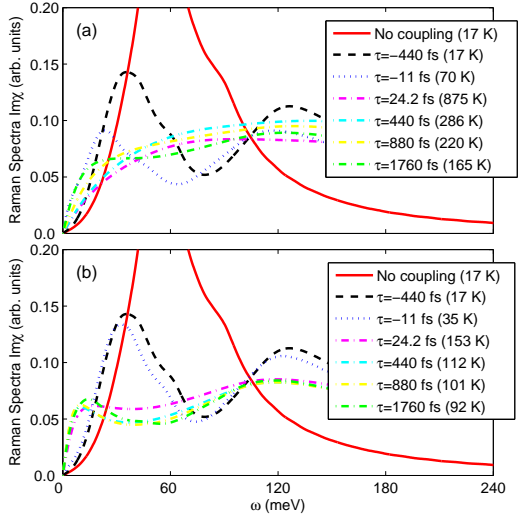}
\caption{(Color online) Time-resolved Raman scattering spectrum
at several different time delays  for (a) $P_0=0.436\;\mu\text{W}$ and (b) $0.022\;\mu \text{W}$. 
The number in parentheses is $T_e$.}
\label{raman}
\end{figure}

\section{Time-resolved Raman scattering spectrum}
\label{sec:chi} 

The time-resolved Raman scattering intensity is calculated via a simple 
relation~\cite{MBakr:2009} from the imaginary part of the Raman response 
function. In the bare vertex approximation,~\cite{TPDevereaux:1994,TPDevereaux:2007} it
is found as:
\begin{equation}
\label{chi}
\text{Im} \chi(i\Omega_m \rightarrow \omega+i\delta) = -\frac{2\pi^2}{N_L}
\sum_\kv \gamma_\kv^2 I_\chi(\kv,\omega)\;, 
\end{equation}
where
\begin{equation}
\label{integral}
I_\chi(\kv,\omega)  = 2i \int d\tau^{\prime}~ \text{Tr} \text{Im} [
\hat{\tau}_{3} {\hat F}^*(\kv,\tau^{\prime})\hat{\tau}_{3}{\hat A}(\kv,\tau^{\prime})]
e^{i\omega \tau^{\prime}}.
\end{equation}
Here
${\hat \gamma}_\kv$ is the nonresonant bare Raman vertex 
given by ${\hat \gamma}_\kv = \gamma_\kv {\hat \tau}_3$ with
${\hat \tau}_3$ being the Pauli matrix and  
$\gamma_\kv = \sum_{\alpha,\beta}e_\alpha^S \frac{\partial^2 \xi_\kv}
{\partial k_\alpha \partial k_\beta} e_\beta^I$.
$\ev^{I,S}$ are the polarization unit vectors of the incident and scattered 
photons, and $\xi_\kv$ the electronic normal-state dispersion of the conduction band. 

Figure~\ref{raman} shows the time evolution of the Raman scattering 
spectrum.  When the electron-hot phonon coupling is switched off, 
the Raman spectrum rises with $\omega$ 
and has a large peak at twice the gap, $2\Delta_0$, at the initial 
temperature (red solid line in Fig.~\ref{raman}). 
A small shoulder in the curve around 90 meV arises due to the van Hove singularity. 
In the presence of the electron-hot phonon 
coupling, the superconducting gap function is renormalized, which shifts the original 
$2\Delta_{0}$-peak to lower frequencies. 
Simultaneously, another peak develops at $2\Delta_0 + \Omega_{ph,1}$ but no peak develops at 
$\Omega_{ph,2}$, for the  same reason for the optical conductivity (Fig.~\ref{sigma}).
After photoexcitation, this double-peak structure evolves into a very broad peak as the system enters the normal state. For large $P_0$, this broad peak remains for a few picoseconds. However, for intermediate $P_0$, once 
the superconducting state recovers, the double-peak structure appears again. This result is fully consistent with our calculations of the TR optical conductivity described above.

\section{Conclusion}
\label{sec:conclusion} 

We have presented a theory for the time-resolved optical conductivity and Raman spectrum, based on the TR spectral function that we have recently formulated for HTSCs. 
Our calculations show that the signature of the electron-$B_{1g}$ mode coupling 
in the TR optical conductivity and Raman spectrum 
is more pronounced than the consequence of the coupling between electrons and the half breathing mode.  
This is the result of a concurrence of anisotropy of the el-ph coupling, 
band structure, and $d$-wave energy gap in HTSCs. Even more interestingly, this 
signature also shows a re-entrant behavior in concurrence with the superconducting 
condensate, which can be controlled by the pump fluence. 
The observation of the broad peak in the TR Raman spectra and their re-entrant 
behavior provides a direct evidence of the el-ph coupling.

\begin{acknowledgments}
One of us (J.-X.Z.) thanks 
A.V. Chubukov, 
G. L. Dakovski, 
T. Durakiewicz, 
F. Marsiglio, 
and G. Rodriguez for helpful discussions.
This work was supported by the National Nuclear 
Security Administration of the U.S. DOE  at  LANL under Contract 
No. DE-AC52-06NA25396, the U.S. DOE 
Office of Basic Energy Sciences, and the LDRD Program at LANL.
\end{acknowledgments}


\begin{thebibliography}{100}

\bibitem{ALanzara:2001} A. Lanzara,
P. V. Bogdanov, X. J. Zhou, S. A. Kellar, D. L. Feng, E. D. Lu, T. Yoshida, H. Eisaki, A. Fujimori, K. Kishio, J. -I.  Shimoyama, T. Noda, S. Uchida, Z. Hussain and Z. X. Shen, Nature (London) {\bf 412}, 510 (2004).

\bibitem{TCuk:2004} T. Cuk, 
F. Baumberger, D. H. Lu, N. Ingle, X. J. Zhou, H. Eisaki, N. Kaneko, Z. Hussain, T. P. Devereaux, N. Nagaosa, and Z.-X. Shen, 
Phys. Rev. Lett. {\bf 93}, 117003 (2004).

\bibitem{TPDevereaux:2004} T. P. Devereaux, 
T. Cuk, Z.-X. Shen, and N. Nagaosa, 
Phys. Rev. Lett. {\bf 93}, 117004 (2004).

\bibitem{GHGweon:2004} G. H. Gweon,
T. Sasagawa, S. Y. Zhou, J. Graf, H. Takagi, D. H. Lee, and A. Lanzara, Nature (Londo) {\bf 430}, 187 (2004).

\bibitem{WMeevasana:2006} W. Meevasana, 
N. J. C. Ingle, D. H. Lu, J. R. Shi, F. Baumberger, K. M. Shen, W. S. Lee, T. Cuk, H. Eisaki, T. P. Devereaux, N. Nagaosa, J. Zaanen, and Z.-X. Shen,
Phys. Rev. Lett. {\bf 96}, 157003 (2006).

\bibitem{DReznik:2006} D. Reznik,
L. Pintschovius, M. Ito, S. Iikubo,  M. Sato, H. Goka, M. Fujita, K. Yamada, G. D. Gu, and J. M. Tranquada, Nature (London) {\bf 440}, 1170 (2006).

\bibitem{JLee:2006} J. Lee,
K. Fujita, K. McElroy, J. A. Slezak, M. Wang, Y. Aiura, H. Bando, M. Ishikado, T. Masui, J.-X. Zhu, A. V. Balatsky, H. Eisaki, S. Uchida, and J. C. Davis, Nature (London) {\bf 442}, 546 (2006).

\bibitem{JXZhu:2006a} J.-X. Zhu, 
 K. McElroy, J. Lee, T. P. Devereaux, Qimiao Si, J. C. Davis, and A. V. Balatsky,
Phys. Rev. Lett. {\bf 97}, 177001 (2006).

\bibitem{MOpel:1999} M. Opel, R. Hackl,
T. P. Devereaux,  A. Virosztek, A. Zawadowski, A. Erb, E. Walker, H. Berger, and 
L. Forr\'{o},   Phys. Rev. B {\bf 60}, 9836 (1999).


\bibitem{JDemsar:1999} J. Demsar,
B. Podobnik, V. V. Kabanov, Th. Wolf, and D. Mihailovic,
 Phys. Rev. Lett. {\bf 82}, 4918 (1999).

\bibitem{PKusar:2008} P. Kusar,
 V. V. Kabanov, J. Demsar, T. Mertelj, S. Sugai, and D. Mihailovic, 
 Phys. Rev. Lett. {\bf 101}, 227001 (2008).

\bibitem{RDAveritt:2002}
R. D. Averitt and A.J. Taylor, J. Phys.: Condens. Matter {\bf 14}, R1357 (2002).

\bibitem{EMChia:2011} E. E. M. Chia, 
J.-X. Zhu, D. Talbayev, and A. J. Taylor, 
Physica Status Solidi RRL {\bf 5}, 1 (2011).

\bibitem{RAKaindl:2000} R. A. Kaindl, 
M. Woerner, T. Elsaesser, D. C. Smith, J. F. Ryan, G. A. Farnan, M. P. McCurry, D. G. Walmsley, Science {\bf 287}, 470 (2000).

\bibitem{RPSaichu:2010} R. P. Saichu,
I. Mahns, A. Goos, S. Binder, P. May, S. G. Singer, B. Schulz, A. Rusydi, 
J. Unterhinninghofen, D. Manske, P. Guptasarma, M. S. Williamsen, and M. R\"{u}bhausen,
 Phys. Rev. Lett. {\bf 102}, 177004 (2009).

\bibitem{LPerfetti:2007}
L. Perfetti, P. A. Loukakos, M. Lisowski, U. Bovensiepen, H. Eisaki, and M. Wolf,
Phys. Rev. Lett. {\bf 99}, 197001 (2007).

\bibitem{NGedik:2005} N. Gedik, M. Langner, J. Orenstein, 
S. Ono, Yasushi Abe, and Yoichi Ando, Phys. Rev. Lett. {\bf 95}, 117005 (2005).

\bibitem{FCarbone:2008} F. Carbone, 
Ding-Shyue Yang, Enrico Giannini, and Ahmed H. Zewail
Proc. Natl. Acad. Sci. {\bf 105}, 20161 (2008).

\bibitem{APashkin:2010} A. Pashkin,
M. Porer, M. Beyer, K. W. Kim, A. Dubroka, C. Bernhard, X. Yao, Y. Dagan, R. Hackl, A. Erb, J. Demsar, R. Huber, and A. Leitenstorfer, Phys. Rev. Lett. {\bf 105}, 067001 (2010).

\bibitem{EJNicol:2003} E. J. Nicol and J. P. Carbotte, Phys. Rev. B {\bf 67}, 214506 (2003).

\bibitem{VVKabanov:1999} V. V. Kabanov, 
J. Demsar, B. Podobnik, and D. Mihailovic, 
Phys. Rev. B {\bf 59}, 1497 (1999).

\bibitem{ARothwarf:1967} A. Rothwarf and B. N. Taylor, Phys. Rev. Lett. {\bf 19}, 27 (1967).

\bibitem{JUnterh:2008} J. Unterhinninghofen, D. Manske, and A. Knorr, Phys. Rev. B {\bf 77}, 
180509(R) (2008). 

\bibitem{CGiannetti:2009} Claudio. Giannetti,
Giacomo Coslovich, Federico Cilento, Gabriele Ferrini, Hiroshi Eisaki, Nobuhisa Kaneko, Martin Greven, and Fulvio Parmigiani,
Phys. Rev. B {\bf 79}, 224502 (2009).

\bibitem{MBeyer:2011} M. Beyer,
D. St\"{a}dter, M. Beck, H. Sch\"{a}fer, V. V. Kabanov, G. Logvenov, I. Bozovic, G. Koren, and J. Demsar, Phys. Rev. B {\bf 83}, 214515 (2011).

\bibitem{GLDakovski:2011} G. L. Dakovski {\em et al.} (unpublished). 

\bibitem{PBAllen:1987}
P. B. Allen, Phys. Rev. Lett. {\bf 59}, 1460 (1987).

\bibitem{JTao:2010}
J. Tao and J.-X. Zhu, Phys. Rev. B {\bf 81}, 224506 (2010).

\bibitem{MRNorman:1995}
M.R. Norman,
M. Randeria, H. Ding, and J. C. Campuzano,    
Phys. Rev. B {\bf 52}, 615 (1995).

\bibitem{FGross:1986}
F. Gross, B.S. Chandrasekhar, D. Einzel, K. Andres, P.J. Hirschfeld, 
H.R. Ott, J. Beuers, Z. Fisk, and J.L. Smith,
Z. Phys. B {\bf 64}, 175 (1986).

\bibitem{JXZhu:2006b}
J.-X. Zhu,
 A.V. Balatsky, T.P. Devereaux, Q. Si, J. Lee, K. McElroy, and J.C. Davis,      
Phys. Rev. B {\bf 73}, 014511 (2006).

\bibitem{FMarsiglio:1991} F. Marsiglio, Phys. Rev. B {\bf 44}, 5373 (1991).

\bibitem{SMaiti:2010} S. Maiti and A. V. Chubukov, Phys. Rev. B {\bf 81}, 245111 (2010).

\bibitem{MBakr:2009}
M. Bakr,  A. P. Schnyder, L. Klam, D. Manske, C. T. Lin, B. Keimer, M. Cardona, and C. Ulrich,
Phys. Rev. B {\bf 80}, 064505 (2009).

\bibitem{TPDevereaux:1994}
T.P. Devereaux,  D. Einzel, B. Stadlober, R. Hackl, D. H. Leach, and J. J. Neumeier, 
Phys. Rev. Lett. {\bf 72}, 396 (1994).

\bibitem{TPDevereaux:2007}
T.P. Devereaux and R. Hackl, Rev. Mod. Phys. {\bf 79}, 175 (2007).

\end{thebibliography}
\end{document}